\documentclass[preprinte, article,publish,moreauthorsq]{mdpi}
\UseRawInputEncoding
\usepackage{amsmath}
\usepackage{amsfonts}
\usepackage{amssymb,bm}
\usepackage{siunitx}
\usepackage{color}
\usepackage{braket}

\firstpage{981}
\pubvolume{4}
\issuenum{3}
\articlenumber{4030065}
\pubyear{2022}
\copyrightyear{2022}
\datereceived{28 July 2022}
\dateaccepted{16 August 2022}
\datepublished{31 August 2022}
\hreflink{https://doi.org/103390/physics4030065} 

\Title{Centenary of Alexander Friedmann's Prediction of the Universe
Expansion and the Quantum Vacuum}

\TitleCitation{Centenary of Alexander Friedmann's Prediction of the Universe
Expansion and the Quantum Vacuum}

\Author{{Galina L. Klimchitskaya} $^{1,2}$\orcidA{} and
Vladimir M. Mostepanenko $^{1,2,3,}$*\orcidB{}}

\AuthorNames{Galina L. Klimchitskaya, Vladimir M. Mostepanenko}

\AuthorCitation{Klimchitskaya, G.L.; Mostepanenko, V.M.}

\address{%
$^{1}$ \quad Central Astronomical Observatory at Pulkovo of the Russian Academy of Sciences, \mbox{196140 Saint Petersburg,  Russia}; g.klimchitskaya@gmail.com\\
$^{2}$ \quad  {Peter the Great Saint Petersburg
Polytechnic University},
195251 Saint Petersburg,   Russia\\
$^{3}$ \quad {Kazan Federal University},
420008 Kazan, Russia}

\corres{Correspondence: vmostepa@gmail.com}

\abstract{{We}
review the main scientific pictures of the universe developed
from ancient times to Albert Einstein and underline that all of them treated
the universe as a stationary system with unchanged physical properties.
In contrast to this, 100 years ago Alexander Friedmann predicted
that the universe expands starting from the point of infinitely large
energy density. {We} briefly discuss the physical meaning of this prediction
and its experimental confirmation consisting of the discovery of redshift
in the spectra of remote galaxies and relic radiation. After mentioning the
horizon problem in the theory of the hot universe, the inflationary model is
considered in connection with the concept of quantum vacuum as an alternative
to the inflaton field. The accelerated expansion of the universe is discussed
as powered by the cosmological constant originating from the quantum vacuum.
The conclusion is made that since Alexander Friedmann's prediction of the
universe expansion radically altered our picture of the world in comparison
with the previous epochs, his name should be put on a par with the names
of Ptolemy and Copernicus.}


\begin{document}
\section{Introduction}\label{Intro}

According to Immanuel Kant \cite{1}, the starry heavens is one of two things
which ``fill the mind with ever new and increasing admiration and awe...``
Questions about what our universe is, how it was created and how long it will
exist have always aroused great interest. In the pre-scientific era,
answers to these questions were usually given on the basis of various myths,
religions, and philosophical systems. In ancient Greece, for the
first time in the history of mankind, the foundations of a scientific approach
to the study and attempts to answer these questions were laid.

In this review, {we} briefly list the
{main}
scientific pictures of the {universe,}
developed during the period of time from ancient Greece to Albert {Einstein,}
and emphasize one characteristic feature common to all of them.
This common feature is that the universe has always been thought as a stationary
system. The concept of a stationary universe was questioned only 100
years ago in 1922 when Alexander Friedmann, on the basis of the general theory
of relativity, demonstrated that our universe expands with time.
This prediction was confirmed experimentally very soon and became the basis
of modern cosmology given every reason to include the name of Alexander
Friedmann on a par with the names of
greatest scientists who completely revised our understanding of the world
around us.

Another important point is that the Friedmann universe exists for a finite
time after its creation at a point ``from nothing''. This initial state of
the universe called the {``cosmological singularity''}
makes a link between the
Friedmann discovery of expanding universe and the concept of quantum
vacuum. According to modern views, the first moments of the evolution of
the universe were governed by quantum theory.
In the framework of a semiclassical {model,} where the gravitational field remains
classical but the fields of matter are quantum, it is possible to consider
the stress-energy tensor of the vacuum of quantized
fields as the source of gravitational field.
Under the influence of quantum vacuum the universe expands exponentially
fast which is known as the cosmic inflation. At slightly later time, inflation gives
way to the Friedmann expansion following the power law. At the present
stage, an expansion of the Friedmann universe is accelerating under the
impact of dark energy. One of the most popular explanations of this
mysterious substance is again given by the quantum vacuum which leads to
a nonzero cosmological constant.

The paper is organized as follows. After a discussion of various pictures
of the universe from ancient times to Einstein in Section~\ref{hystory},
{we} briefly consider the main facts of Alexander Friedmann's
scientific biography in
Section~\ref{Friedmann}. Section~\ref{predict} is devoted to  Friedmann's
prediction of the universe expansion made 100 years ago.
The experimental facts confirming that the universe is really expanding are
presented in Section~\ref{expt}. Section~\ref{birth} is
devoted to the cosmic inflation and creation of the universe from the quantum vacuum.
In Section~\ref{accelerate}, {we} consider the accelerated expansion of the
universe and its explanation in terms of dark energy originating from the
vacuum of quantized fields.
In Section~\ref{discuss}, the reader will find the discussion,
and {the paper ends} with conclusions in Section~\ref{concl}.

\section{Pictures of the Universe---From Ancient Times to Albert
Einstein}\label{hystory}

The first scientific picture of the universe based on observations was
created by Claudius Ptolemy in the first century AD.
The Ptolemy system was geocentric which means that the Earth was placed
at the center of the world. All remaining celestial bodies, i.e., the
Moon, the Sun and five planets known at that time (Mercury, Venus, Mars,
Jupiter, and Saturn), rotated around the Earth in circular orbits.
According to Ptolemy's system of the world, beyond Saturn there is a
firmament to which the fixed stars are attached. It was assumed that
the stars and the firmament do not obey the same physical laws as all
bodies on the Earth. Despite the presence of some nonscientific
elements, the Ptolemy system gave the possibility to perform calculation
of both future and past positions of the Moon, Sun and all five planets
with rather high accuracy.
{In fact,}
 this system was successfully used
until the 16th century. Needless to say, Ptolemy's picture of the
universe was stationary. It did not vary with time.

Important change in our picture of the world has been made by investigations
of Nicolaus Copernicus, Johannes Kepler, and Galileo Galilei, performed in
the 16th and 17th centuries. They developed on a scientific basis and
supported by observations the long-proposed hypothesis that our
universe is in fact heliocentric. According to the heliocentric system,
the Earth and all five planets orbit the Sun whereas the Moon orbits the
Earth. Johannes Kepler made an important discovery that the orbits of planets
are not circles but ellipses with the Sun at one of the ellipse's focuses.
Both Copernicus and Kepler believed Ptolemy's idea that the stars are
fixed points attached to the firmament. However, Galilei elaborated methods
to determine the shape of stars and made estimations of their radii.
The radically new picture of the world established by Copernicus, Kepler,
and Galilei retains its validity in the scales of Solar system up to the
present. However, they persisted in the belief that the universe is stationary
in a sense that planets followed and will always follow the same
predetermined orbits.

The next dramatic step in our understanding of the universe was made by
Isaac Newton who developed the first physical theory, Newtonian mechanics,
and laid foundations of the mechanical picture of the world. In his book
\cite{2}
{\it {Mathematical Principles of Natural Philosophy}}
published in
1687, Newton formulated the three laws of mechanics and the law of gravitation
which must be obeyed by all material bodies on the Earth and in the sky.
Newton arrived to the fundamental conclusion that the inertial {mass, $m_i$,}
of each material {body}
is equal to its gravitational {mass, $m_g$,} which is responsible
for the gravitational attraction. This was the first formulation of the
equivalence {principle} used by Einstein as the basis of general relativity
theory 230 years later.

With the second law of mechanics and the Newton law of gravitation, one finds
that the force acting between the test mass $m_i=m_g$ and the Earth of the mass
$M$ and radius $R$ can be expressed in two {ways:}
\begin{equation}
F=m_ia=\frac{Gm_gM}{R^2},
\label{eq1}
\end{equation}
\noindent
where $G$ is the gravitational constant and $a$ is the acceleration of the test
mass. Then, using the equivalence principle, {one} obtains
from Equation (\ref{eq1}):
\begin{equation}
a=\frac{GM}{R^2},
\label{eq2}
\end{equation}
\noindent
i.e., the conclusion is that all bodies in the vicinity of Earth surface fall down
with the same acceleration independently of their mass. The law that light
and heavy bodies falling to the ground from the same height reach the ground
at the same time was experimentally discovered by Galileo Galilei. Newton
derived it theoretically. This fundamental law of Nature, which defies common
sense, was destined to play a huge role in elucidating the structure and
evolution of our universe.

Newton's concept of the universe pushed its boundaries far beyond the
solar system. According to Newton, the universe is infinitely large in volume
and contains infinitely many stars. The space of the universe is homogeneous
(i.e., all points are equivalent) and isotropic (i.e., all directions are also
equivalent). However, keeping unchanged an important element of the previous
pictures of the universe, Newton believed that our universe is of an infinitely large
age and it will exist forever. In this sense he considered the universe to be
stationary.

Newton's picture of the universe was universally accepted until the early
20th century despite some unresolved problems. For instance, according to
Olbers paradox proposed in 1823, in the case of an infinitely large universe,
in every direction one looks, one should see a star. As a result,  the entire night sky
 would shine like a surface of a star, which is not true.
One more difficulty is the problem of the heat death of the universe discussed
by
Bailly in 1777 and elaborated on by Lord Kelvin in 1851 on the basis of the laws
of thermodynamics. Since the Newtonian universe exists for an infinitely large time,
it should already have reached a state where all energy is evenly distributed and all
dynamical processes are terminated. Thus, the observed temperature differences are
in contradiction to an assumption that the universe is infinitely old.

A new era in the study of the universe began in 1915 when Albert Einstein created
the general theory of relativity starting from the equivalence principle.
According to this theory, there is no gravitational
{force} which attracts material bodies to each other. All bodies move freely along
the shortest (geodesic) lines in the Riemann curved space-time of
the {universe} which
becomes curved under the impact of energy and momentum of these bodies. Thus, the
general theory of relativity  describes the self-consistent {system} where the space-time
curvature is determined by the material bodies whereas their motion is caused by
the character of this curvature.

The description of the universe as a whole in the framework of general theory of
relativity is based on {Einstein}
 field {equations,}
\begin{equation}
R_{ik}-\frac{1}{2}Rg_{ik}-\Lambda g_{ik}=
\frac{8\pi G}{c^4}T_{ik},
\label{eq3}
\end{equation}
\noindent
where $R_{ik}$ is the Ricci tensor characterizing the  space--time curvature,
$R=g^{ik}R_{ik}$ is the scalar curvature, $g_{ik}$ is the metrical  tensor,
$\Lambda$ is the cosmological constant, $c$ is the speed of light, and
$T_{ik}$ is the stress-energy tensor of matter. The indices $i$ and $k$
here take the values 0,\,1,\,2,\,3 and there is a summation over the
repeated indices.

The term $\Lambda g_{ik}$ was absent in the original {Einstein's}
{equations,} published
in 1915 \cite{3} {(see} \cite{3e} for English translation).
Einstein introduced it when applying his field equations to the
universe as a whole in order to compensate the effect of gravity and make the
universe {stationary} \cite{4} {(see} \cite{4e} for English translation).
This means that he shared the opinion of Ptolemy,
Copernicus, Galilei and Newton that the universe does not vary with time.
Using the basic concepts of Newton's picture of the world, Einstein also assumed
that the 3-dimensional space of the universe is homogeneous and isotropic.
Based on this assumptions, he obtained the model of the stationary universe of finite
spatial volume. This universe exists forever. It has never been created. There
is a finite number of stars in the Einstein universe.

Thus, all the greatest scientists
from Ptolemy to Einstein, who determined the views of
 mankind on the universe for two millennia believed that it is stationary.
A new era in the understanding of the universe began with the paper by Alexander
Friedmann \cite{5}, published 100 years ago, in 1922, in which he first
proved that the universe expands. Before considering Friedmann's discovery,
{we}
briefly present the main facts of his scientific biography, making it clear how
he came to such  a radical conclusion.

\section{Brief Scientific Biography of Alexander Friedmann}\label{Friedmann}

\begin{figure}[H]
\vspace*{-3.8cm}
\includegraphics[width=9.5cm]{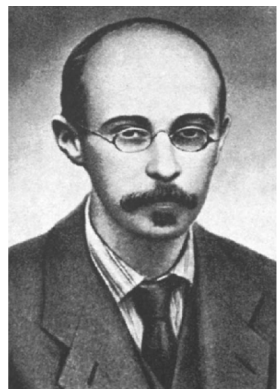}
\vspace*{-3.5cm}
\caption{
{Alexander}
A.~Friedmann (6 June  1988 ---
 16 September 1925), the founder of modern cosmology.
\label{fg1}}
\end{figure}

Alexander Friedmann {(see his photo in
Figure} \ref{fg1})
was born on 6 June, 1888 in Saint Petersburg (the capital of Russian Empire)
in the artistic family
(a full description of his life can be found in \cite{6,7}).
Alexander Friedmann's father's name was also Alexander. He was an artist at the
Court {Ballet}
of the Imperial Theater and a ballet composer.
Alexander Friedman's mother's maiden name was Lyudmila Voyachek, she was a pianist.
She graduated from the Saint Petersburg Conservatoire. Nothing indicated that a
child born in such a family will become {an eminent}
mathematician and physicist.
After the divorce of Friedmann's parents in 1897,  he lived with his father.
In the same year, he started to study at the Second Saint Petersburg High School
which was known for the highly qualified teachers in the field of
mathematics and physics.
While still a schoolboy, Alexander Friedmann, in collaboration with his
schoolmate Yakov Tamarkin
(in the future, {a}
famous mathematician), wrote his first {paper,}
devoted to Bernulli numbers. In 1906, this {paper}
 was published in
 {\it {Math. Ann.}}
by
the recommendation of David Hilbert.

Alexander Friedmann graduated from
{High School}
 in 1906 with a gold medal and was
admitted to the first course of the Department of Mathematics belonging to the
Faculty of Physics and Mathematics at {the}
Saint Petersburg University. During his
student years at the university, Alexander Friedmann obtained an intimate knowledge
in different fields of mathematics and physics and his successes were always
evaluated as ``excellent''.

After a graduation from the Department of Mathematics in 1910, Alexander Friedmann
stayed at the same department as a Postgraduate Researcher and to prepare to the Professor position.
His supervisor was the famous mathematician academician
Vladimir Steklov. From 1910 to 1913, Alexander Friedmann solved
several complicated problems in mathematical physics, published many papers and
delivered lectures in mathematics for students. Although he successfully passed
 examinations for a Master degree, he formally defended the Master thesis
only in 1922. By that time, he  was already Full Professor at {the} Perm University
(1918--1920), at the Petrograd University, Petrograd Polytechnic Institute, and
at the Institute of Railway Engineering (Petrograd, formerly Saint Petersburg,
1920--1925).

In 1913, Alexander Friedmann was employed by the Saint Petersburg Physical (later
renamed in Geophysical) Observatory. During the work at this institution, he
obtained several fundamental results in dynamical meteorology, hydrodynamics,
and aerodynamics. His results retain their significance to the present day, and
Friedmann's name is well
{known}
to everyone working in these fields.
Several important results were obtained by him during the visit to Leipzig
University in the first part of 1914.

During the three years of the World {War I,} from 1914 to 1917, Friedmann served in
the Air {Force}
of the Russian Empire. During this period of his life, he
 personally
piloted airplanes, organized the aerological service, and created the mathematical
theory of bombing. His service during the war was marked with several military
awards.

The period of Alexander Friedmann's life from 1920 to 1925 was especially productive.
During these years, he published several books and obtained outstanding
results in the field of dynamical meteorology. As a recognition of his scientific
merits, in 1925 Alexander Friedmann was appointed Director of the Geophysical
Observatory of the Russian Academy of Sciences. Just during this period of {time,}
he
published two papers \cite{5,8} containing an extraordinary prediction that our
universe expands. {Although} on 16 September  {1925} Alexander Friedmann
tragically died of typhus at the age of 37, these papers made his name immortal.
In the next section, {we} briefly discuss the essence of the obtained results and
how the distinctive features of Friedmann's scientific career helped him make
a discovery that {even} the great
Einstein himself missed.

\section{Friedmann's Prediction of Expanding Universe}\label{predict}

In his approach to the description of the universe in the framework of  general
relativity theory, Friedmann assumed that the three-dimensional space is
homogeneous and isotropic. In this regard, he followed his predecessors Newton
and Einstein. The assumption of homogeneity and isotropy alone gives the
possibility to find the metric, i.e., the {distance}, $ds$,
between two infinitesimally close
space-time {points,} $x^i$ and $x^k$, using the standard mathematical {methods:}
\begin{equation}
ds^2=g_{ik}dx^idx^k=c^2dt^2-a^2(t)[d\chi^2+
f^2(\chi)(d\theta^2+\sin^2\theta d\varphi^2)].
\label{eq4}
\end{equation}

Here, $t$ is the time coordinate whereas $\chi$, $\theta$, and $\varphi$ are analogous
to the spherical coordinates in the three-dimensional space.
{In doing so,}
the usual
Cartesian coordinates and the radial coordinate are expressed as
\begin{eqnarray}
&&
x^1=r(t)\sin\theta\cos\varphi, \qquad
x^2=r(t)\sin\theta\sin\varphi,\qquad
x^3=r(t)\cos\theta,
\nonumber \\
&&
r(t)=a(t)f(\chi).
\label{eq5}
\end{eqnarray}
\noindent
{The} function $f(\chi)=\sin\chi$,
0, and $\sinh\chi$ depending on whether the constant
curvature of the three-space is equal to $\kappa=1$, 0, and
{$-$1},
respectively.

The function $a(t)$ is called the {``scale factor.''}
It has the dimension of length.
In the case of $f(\chi)=\sin\chi$ (the space of positive {curvature),}
$a(t)$ has the
meaning of the radius of curvature. The space of positive curvature has the finite
volume $V=2\pi^2a^3(t)$. The spaces of zero and negative curvature have an infinitely
large volume. In {his}
first paper \cite{5}, Friedmann considered the space of positive
{curvature,} $\kappa=1$, whereas his second paper \cite{8} is devoted to the space of
negative curvature $\kappa=-1$.

In the Friedmann approach to the problem, it is important that he was a {mathematician}
who used the rigorous analytic methods. He wished to see what is contained in the
fundamental {Einstein's}
equations (\ref{eq3}) in the case of a homogeneous isotropic
metric (\ref{eq4}) independently of our historical and methodological preferences.
This approach, which proved to be very fruitful in all Friedmann's diverse scientific
activities, was based on the long-standing traditions of the
Saint Petersburg mathematical school.

Substituting Equation (\ref{eq4}) in Equation (\ref{eq3})
and calculating $R_{ik}$ and $R$ by the
standard expressions of Riemann geometry, Alexander Friedmann obtained two equations,
which were later named after {him:}
\begin{eqnarray}
&&
\frac{d^2a}{dt^2}=-\frac{4\pi G}{3c^2}a(\varepsilon+3P)+
\frac{1}{3}c^2a\Lambda,
\nonumber \\
&&
\left(\frac{da}{dt}\right)^2=\frac{8\pi G}{3c^2}a^2\varepsilon -
\kappa c^2+\frac{1}{3}c^2a^2\Lambda.
\label{eq6}
\end{eqnarray}

{In} these equations, it was taken into account that in the homogeneous isotropic space
the stress-energy {tensor,} $T_{ik}$, is diagonal and that its component
$T_0^{\,0}=\varepsilon$ has the meaning of the energy density of matter, whereas
its {components,} $T_1^{\,1}=T_2^{\,2}=T_3^{\,3}=-P$, describe the pressure $P$ of matter.
Note also
 that in {his}
 papers \cite{5,8}{ Friedmann} considered the so-called
 {``dust matter''} for which the relative velocities of its constituents are small as
compared to the speed of light.
This leads to the zero {pressure, $P=0$,} but does not
affect any of the fundamental conclusions following from Equation (\ref{eq6}).

Friedmann found that for {$\kappa=1$} Equation (\ref{eq6}) admits the
stationary solution in the special case when
\begin{equation}
\varepsilon+3P=\frac{c^4\Lambda}{4\pi G},\qquad
\frac{4\pi Ga^2}{c^4}(\varepsilon+P)=\kappa.
\label{eq7}
\end{equation}
\noindent
{In} this {case, from Equation}
(\ref{eq6}) one {has:}
\begin{equation}
\frac{d^2a}{dt^2}=\frac{da}{dt}=0,\qquad a=a_0={\rm const}.
\label{eq8}
\end{equation}

{This} is the stationary universe obtained earlier by Einstein \cite{4,4e}.
The stationary solution exists only in the case $\kappa=1$, $\Lambda\neq 0$.
If $\Lambda=0$, the first equality in
Equation (\ref{eq6}) leads to $d^2a/dt^2<0$
because for the usual matter it holds $\varepsilon+3P>0$.
For $\kappa=-1$ the stationary universe containing matter with  $\varepsilon>0$
is impossible \cite{8}. Thus, the cosmological {solution} found by Einstein is an exceptional case, whereas in all other cases (i.e., with $\Lambda=0$ or
$\Lambda\neq 0$ but with conditions (\ref{eq7}) {violated),} the universe is
nonstationary so that the scale factor $a(t)$ depends on time.

Friedmann derived one more specific solution of
Equation (\ref{eq6}){ for} which the
scale factor $a$ depends on time but the scalar curvature {$R$} is constant.
This is the {solution} previously found by de Sitter \cite{9}.
It is most simple to illustrate the de Sitter solution, which is determined
by only the cosmological constant in the absence  of usual matter,
$\varepsilon=P=0$, for the case $\kappa=0$. Then Equation (\ref{eq6}) is simplified
to one {equation:}
\begin{equation}
\frac{da}{dt}=\sqrt{\frac{\Lambda}{3}}ca,
\label{eq9}
\end{equation}
\noindent
which has the {solution}
\begin{equation}
a(t)=a_0\,\exp\left(c\sqrt{\frac{\Lambda}{3}}t\right).
\label{eq10}
\end{equation}

The scalar curvature of the homogeneous isotropic spaces is given by
\begin{equation}
R=-\frac{6}{a^2}\left\{\frac{1}{c^2}\left[
\left(\frac{da}{dt}\right)^2+a\frac{d^2a}{dt^2}\right]+\kappa\right\}.
\label{eq11}
\end{equation}

{Substituting Equation} (\ref{eq10}) in {Equation} (\ref{eq11}) for $\kappa=0$, one obtains the
constant scalar curvature of the de Sitter {space--time} $R=-4\Lambda$.
We will return to the consideration of the de Sitter scale factor (\ref{eq10})
in Section~\ref{birth} in connection with the quantum vacuum.

As to the general solution of Friedmann equation (\ref{eq6}), it is characterized
by the zero initial value of the scale factor $a(0)=0$ and by the infinitely large
values of the initial scalar curvature $R(0)$ and energy density $\varepsilon(0)$.
Thus, at the initial moment the space of the universe was compressed to a point
and the time period from the creation of the world to the present moment is
finite \cite{5}. {The initial state of the universe is called the cosmological
singularity;
{see} \cite{10,11} for
details
 about the solutions of Friedmann equation
(\ref{eq6}) for different
{types}
of matter and
{corresponding}
equations of state.

It should be noted that the first reaction of Einstein to Alexander Friedmann's
results was
{completely}
negative.
{Shortly}
after the publication of Friedmann's paper
\cite{5}, Einstein published a note \cite{12}
{(see} English translation in \cite{12e}) claiming that the cosmological
{solutions,} found by {Friedmann,} do not satisfy the field equations of general
relativity theory. In response to this criticism, Friedmann wrote the explanation
{letter} \cite{AAF2AE} (for English translation, see \cite{AAF2AEe}),
which was put in Einstein's hands by
{Yurii}\ A.\ Krutkov during his
visit to Germany.
This letter contained exhaustive {explanations} which
{cannot}
be ignored. As a result, Einstein published one more {note} where he recognized
that his criticism ``was based on an error in calculations''
\cite{13} {(see} \cite{13e} for English transaltion).
Even after Einstein had realized that his original criticism was incorrect,
this did not make {him}
 a supporter of the concept of an expanding universe.
For a long time Friedmann's discovery went largely unnoticed. Its importance
became obvious only after the {publications by}
{Georges}
Lema\^{e}tre and others (see below).

{In fact,}
Alexander Friedmann did not construct a new physical theory for the
description of the universe. This was performed by Albert Einstein, who created  the
general theory of relativity. The {greatness} of Alexander Friedmann lies in the fact
that he was the first to believe that the mathematical solution of Einstein
{equations,} corresponding to an expanding {universe,}
{can}
indeed apply to the real
{world} at the time when this idea was not considered plausible. In doing this,
he changed our picture of the universe and gave start to {grandiose}
 cosmological
investigations of the last century.

Although Friedmann himself believed that the observational data at our disposal
are completely insufficient for choosing the solution of his equations that
describes our universe \cite{5}, the first experimental confirmation of the
universe expansion came very soon.

\section{Experimental Confirmations of the Universe Expansion}\label{expt}

For all nonstationary solutions of the Friedmann equation (\ref{eq6}) distances
between any two remote bodies in the observed universe increase with time.
This is seen from Equation~(\ref{eq4}), {where} the spatial distance is proportional to
the scale factor $a(t)$. As a result, if the universe expands, all observable
galaxies should move away from the Earth. The galaxies are observed due
to {the light}
emitted by them. According to the Doppler law, the frequency of an electromagnetic
{wave} emitted by a source moving away from the {observer} is decreased. This is the
so-called {``redsqhift''} of the emitted light to the red end of the visible~spectrum.

Actually,
the first observation of the redshift of Andromeda Nebula was made by
Slipher in 1913 \cite{14}, i.e., before the development of the
general theory of relativity
by Einstein. He interpreted this observation in the spirit of Doppler effect that
the Andromeda Nebula moves away from the Earth. The universal {law which connects}
the
redshift in the spectra of remote galaxies with the universe {expansion} was
experimentally discovered in 1927 by Georges Lema\^{e}tre \cite{14a} and in 1929 by
Edwin Hubble \cite{14b} {who finally} validated that the nebulas are the galaxies outside
the Milky Way. According to this law, the velocity of a remote galaxy is proportional
to a distance to it, $v=HD$, where $H$ is the Hubble {constant}
 which can be expressed
via the scale factor as
\begin{equation}
H=H(t)=\frac{1}{a(t)}\,\frac{da(t)}{dt},
\label{eq12}
\end{equation}
\noindent
i.e., it is {in fact}
time-dependent.

The discovery of the redshifts in the spectra of remote galaxies was the first
experimental confirmation of the universe {expansion} predicted by Alexander Friedmann.
The next important step was made by George {Gamow}
who elaborated the theory of  a hot
{universe} which
{provided a}
possibility to explain the creation of chemical elements
and the formation of galaxies \cite{15}. The basic point of {Gamow's}
 theory was
an assumption that the early universe was dominated not by the dust matter but by
radiation with the equation of {state} $P(t)=\varepsilon(t)/3$. In the framework of
{Gamow's}
theory of hot {universe,} Ralph Alpher and Robert Herman predicted the
existence of the background relic radiation \cite{16}. The discovery of this radiation
served as the second most important experimental confirmation of the expansion
of the universe.

The cosmic microwave background electromagnetic radiation,
{called also the
``relic
radiation,''} was discovered in 1965 by Arno Penzias and Robert Wilson \cite{17}.
It fills all space and was created in the epoch of formation of first atoms.
The observation of relic radiation confirmed the origin of the universe from the
cosmological {singularity} predicted by Alexander Friedmann as a result of the
so-called {``Big Bang.''}
Based on the theory of the hot universe, it became possible to
describe the various eras in the universe evolution starting from the
Electroweak Era followed
 by the Particle Era, the Era of Nucleosynthesis, Eras of Nuclei, Atoms,
and, finally, by the Era of Galaxies.
This covers the period of the universe
evolution from approximately $10^{-33}~$s after
the cosmological singularity to
about 14 billion {years} which is the present age of the universe. As to the very early
stage from 0 to $10^{-33}~$s, it remained a mystery and could not be explained
on the basis of the general theory of relativity.

\section{Cosmic Inflation and Creation of the Universe from Quantum Vacuum}\label{birth}

As was noted above, the basic
{assumption,} used in Friedmann's cosmology and in the
theory of hot {universe,} is that the space is homogeneous and isotropic. This assumption
was confirmed by the approximately homogeneous and isotropic large-scale distribution
of galaxies and, more importantly, by the properties of relic radiation.
It was found that the relic radiation has a blackbody thermal spectrum at
$T=2.726\pm 0.001~$K average temperature and the variations of this {temperature} measured
from different directions in the {sky} do not exceed $\Delta T/T\sim 10^{-5}$.

As discussed in Section~\ref{expt}, at the early stages of its evolution the
universe {was}
filled with radiation possessing the equation of {state} $P=\varepsilon/3$.
In this case the solution of {Friedmann}
 equation (\ref{eq6}) is given by
$a(t)\sim\sqrt{t}$ and the respective energy density behaves as
$\varepsilon(t)\sim 1/t^2$ when $t$ goes to zero.

These behaviors, however, create a problem. The {point}
is that if
$t$ is decreased down
to the Planck {time} defined as
\begin{equation}
t_{\rm Pl}=\sqrt{\frac{\hbar G}{c^5}}=5.39\times 10^{-44}~\mbox{s},
\label{eq13}
\end{equation}
\noindent
the size of the universe turned out to be unexpectedly large
$a(t_{\rm Pl})\sim 10^{-3}~$cm as compared to the Planck {length,}
\begin{equation}
l_{\rm Pl}=t_{\rm Pl}c=\sqrt{\frac{\hbar G}{c^3}}=1.62\times 10^{-33}~\mbox{cm}\, ,
\label{eq14}
\end{equation}
\noindent
traveled by light during the Planck time.

Thus, if the above scale factor were applicable down to $t=0$, at Planck time the
universe would consist of the $10^{89}$ causally disconnected parts. This is in
contradiction with the fact that the relic radiation in all places and all directions
in the sky has the same temperature {--- the}
 so-called {``horizon problem.''}

The horizon problem cannot be solved in the framework of the  general
theory of relativity.
The point is that this is the classical theory and the space-time scales of the order
of Planck length and Planck time are outside the region of its applicability.
In the absence of quantum theory of gravitation, which is still unavailable in spite
or repeated attempts to develop it undertaken during several decades, some
semiclassical approaches are believed to lead to
at least a partial solution of the problem.

In 1981, Alan Guth \cite{18} found that the symmetry {breaking} caused by the
scalar {fields} introduced in particle {physics}
can cause the period of exponentially
fast expansion of the universe. During this period, the scale factor varies as
$\exp(t)$ rather than $\sqrt{t}$. As a result, at the Planck time the universe has
the Planck size which solves the horizon problem. It {was}
{Guth} who introduced
the term {``inflation''} for the exponentially fast expanding universe.
The scalar field
responsible for the inflation process was called the {``inflaton field.''}
The theory of
inflation was further developed by Andrei Linde \cite{19}.
Actually,
the possibility of an early exponential expansion of the universe was
predicted before
Guth by Sergey Mamaev and {one of the authors of this paper}
\cite{24} {and, independently, by}
Alexei Starobinsky \cite{25} based on the semiclassical {Einstein's}
 equations (see
below).
{In doing so,}
 the scale factor was expressed either via the proper
synchronous {time} $t$ \cite{25} or, equivalently, via the conformal {time}
$\eta$ \cite{24}.

Due to a very fast expansion of the universe during the inflationary stage,
the energy density
of matter becomes very low. The conversion of the energy density of oscillating
inflaton field into that of usual matter is called {``reheating''} after inflation.
The theory of reheating was developed by Lev Kofman, Anfrei Linde, and Alexei
Starobinsky \cite{20} using the effect of resonant particle production in the
time-periodic external field revealed earlier by one of the authors and
Valentin Frolov \cite{21}.

The weak point of the theory of inflation is that the physical  nature of inflaton
field remains unclear. In this situation, the question arises of whether there are
other possibilities for obtaining the period of exponentially fast expansion in
the evolution of the early universe.
The possible answer to this question was given by the theory
of quantum matter fields in curved space-time \cite{22,23}. This theory is applicable
under the condition that the gravitational field can be considered as the classical
background, i.e., at $t\gg t_{\rm Pl}$, which is already well satisfied at
$t\geqslant 10^{-40}~$s. One can assume that at $t\sim 10^{-40}~$s all quantum
matter fields are in the vacuum state $|0\rangle$, i.e., the number of particles of
different kinds is zero. This does not mean, however, that the vacuum energy density
and pressure are zero because vacuum is polarized by the external gravitational field.

The vacuum stress-energy tensor of quantum matter fields with different spins in the
homogeneous isotropic space was calculated in the 1970s
 by several groups of {authors.}
It is common knowledge
that the quantities $\langle0|T_{ik}|0\rangle$ contain
the ultraviolet divergences. These divergences can be interpreted in terms of the
bare cosmological constant which is connected with the infinitely large energy
density of the zero-point oscillations, the bare gravitational constant, and the
bare constants in front of the invariant quadratic combinations of the components
of Ricci tensor.
The finite expressions for the renormalized values
of $\langle0|T_{ik}|0\rangle_{\rm ren}$, obtained after the removal of
{divergencies,} can be found
in \cite{22,23}. Based on these results, the so-called {``self-consistent''}
Einstein
equations were {considered:}
\begin{equation}
R_{ik}-\frac{1}{2}Rg_{ik}=\frac{8\pi G}{c^4}\langle0|T_{ik}|0\rangle_{\rm ren}.
\label{eq15}
\end{equation}

In these equations, the vacuum of quantized matter fields $|0\rangle$ is
polarized by the gravitational field of the homogeneous isotropic space with metric
(\ref{eq4}) determining the left-hand side of {Equation} (\ref{eq15}).
In free Minkowski space-time, the physical energy density and pressure in the
vacuum state obtained after discarding of infinities are equal to zero.
However, in an external field (regardless of whether it is electromagnetic or
gravitational) after discarding of infinities the vacuum state is polarized,
such as a dielectric in an electric field, i.e., it is characterized by some nonzero
stress-energy tensor.
On the other hand, the gravitational field described by the left-hand side of {Equation}
(\ref {eq15}) is created as a source by the vacuum energy density and pressure
on the
right-hand side. By solving the self-consistent {equations}
(\ref{eq15}),
one can
find the scale factor of the homogeneous isotropic space determined by the quantum
vacuum of the matter fields.

It should be stressed that the theoretical {approach} based on Equation (\ref{eq15})
is semiclassical. This means that the gravitational field and the corresponding
metrical tensor in Equation (\ref{eq4}) are still treated as the classical ones.
It is assumed that only the fields of matter (scalar, spinor, vector, etc.)
exhibit a quantum behavior.
Recall that this approach is applicable at
$t\geqslant 10^{-40}~$s. At earlier moments down to the Planck time and to the
domain of singularity in the solution of the classical general theory of relativity,
one should take into account the effects of quantization of space-time, i.e.,
the effects of quantum gravity.

The solutions of Equation (\ref{eq15}) for the massless matter
fields were obtained in \cite{24,25}.
As an example, for a scalar field in the space of positive
curvature the self-consistent scale factor is given by
\begin{equation}
a(t)=\sqrt{\frac{\hbar G}{360\pi c^3}}
\cosh\left(t\sqrt{\frac{360\pi c^5}{\hbar G}}\right).
\label{eq16}
\end{equation}

{Using} Equations (\ref{eq13}) and (\ref{eq14}),
{one can see}
{that for $t>t_{\rm Pl}$}
the scale factor (\ref{eq16}) takes the form
\begin{equation}
a(t)=\frac{l_{\rm Pl}}{\sqrt{360\pi}}
\exp\left(\frac{\sqrt{360\pi}\,t}{t_{\rm Pl}}\right), 
\label{eq17}
\end{equation}
\noindent
i.e., it is the exponentially increasing with time scale factor of the de Sitter
space ({c.f.}  with
Equation (\ref{eq10})).
This is the scale factor describing the cosmic inflation obtained
on the fundamental grounds
of quantum field theory without introducing the inflaton field.

 Actually,
in this approach the inflationary universe is spontaneously created from
the quantum vacuum. It should be noted that the expressions for
$\langle0|T_{ik}|0\rangle_{\rm ren}$ contain the third and fourth derivatives of the
scale factor, which lead to scalar and tensor instabilities.   As a result,
the de Sitter solution becomes unstable relative to the spatially homogeneous massive
scalar modes (scalarons). Using this fact, Alexei Starobinsky \cite{25} constructed
the  nonsingular cosmological model where the de Sitter universe describing
inflation is spontaneously created from the quantum vacuum. Then, due to the generation
of scalarons and
 {the
decay of scalarons}
into usual particles, the exponentially fast expansion
of the inflationary stage is replaced by the power-type expansion {law}
$a(t)\sim\sqrt{t}$ of the theory of hot universe. According to current concepts,
the inflationary stage of the universe expansion lasts from
$t\sim 10^{-36}~$s to $\sim 10^{-33}~$s. More precise measurements of the spectrum
of relic {radiation,} planned
 {for the}
near {future,} should provide information concerning the
gravitational radiation generated during the exponentially fast expansion.
This will help to conclusively establish the main properties of the
inflationary stage of the universe evolution.

\section{Accelerated Expansion of the Universe, Dark Energy and
the Quantum Vacuum}\label{accelerate}

According to the commonly accepted {views} formed by the end of the 20th century,
the present stage of the universe evolution is
 described by the Friedmann equations
(\ref{eq6}) with $\Lambda=0$ and a predominance of dust matter having the energy
{density} $\varepsilon(t)=\rho(t)c^2$, where $\rho(t)$ is the mass of matter per
unit volume, and zero {pressure} $P=0$. In doing so,
the fraction of visible matter is by a factor of 5.4 smaller than {that}
of invisible {matter, which}
possesses the same properties as the visible one and
is called the {``dark matter.''}
We know about the existence of dark matter due to its
gravitational action on visible bodies.
In this situation, it was expected that due to
the gravitational attraction of matter the expansion of the universe should be
decelerating.

It was quite unexpected, however, when in 1998 {two}
research teams (Supernova
Cosmology Project and High-Z Supernova Search Team) observed clear evidence that,
{by}
contrast, the expansion of the universe is {accelerating (see the references
and discussion below).}
This
means that the
velocities of remote galaxies moving away from the Earth increase with time.
Actually,
the inflationary stage of the universe evolution was also the period of
accelerated expansion which, however,  lasted for an infinitesimally short period
of time. As to the accelerated expansion observed at present, it has already been
going on
for  several billion years.

The {matter} which causes acceleration of the universe {expansion} was called
{``dark energy.''} Unlike the dark matter, which is not seen but gravitates like
ordinary matter, dark energy acts against the gravitational attraction,
i.e., it is characterized by the negative pressure.
The observed acceleration rate of the universe expansion requires that 68\% of
the universe should consist of dark energy. As to
dark matter and
usual
visible matter, they constitute approximately 27\% and
5\% of the universe's
energy, {respectively}.
 There were many attempts in the literature to understand the physical
nature of dark energy by introducing some new hypothetical particles with
unusual properties \cite{26}. However, the most popular explanation of the
accelerated expansion returns us back to the concept of cosmological constant and
quantum vacuum.

As discussed in Section~\ref{predict}, in the absence of background matter,
$\varepsilon=P=0$, the cosmological {term $\Lambda g_{ik}$} in the {Einstein's}
 equations
(\ref{eq3}) determines the de Sitter space-time with a scale factor (\ref{eq10}).
In the homogeneous isotropic space, the presence of
this term just corresponds to the energy
density and {pressure,}
\begin{equation}
\varepsilon_{\Lambda}=\frac{c^4\Lambda}{8\pi G}>0,
\qquad
P_{\Lambda}=-\frac{c^4\Lambda}{8\pi G}=-\varepsilon_{\Lambda},
\label{eq18}
\end{equation}
\noindent
i.e., results in some effective negative pressure. When the cosmological term
is considered along with the stress-energy tensor of the ordinary matter
$T_{ik}$, it just leads to the required acceleration of the universe expansion.
{Calculations}
show that an agreement with the observed rate of acceleration is
reached for $\Lambda\approx 2\times 10^{-52}~\mbox{m}^{-2}$ and
{the corresponding}
energy {density} $\varepsilon_{\Lambda}\approx 10^{-9}~\mbox{J/m}^3$
\cite{30,31}.

The question arises  what is the nature of the cosmological constant.
 Actually,
it can be considered as one more fundamental constant closely
{related to}
the concept of the quantum vacuum \cite{27}. This statement is based on the
geometric structure of the vacuum stress-energy tensor of quantized {fields,}
\begin{equation}
\langle 0|T_{ik}(x)|0\rangle=Ig_{ik},
\label{eq19}
\end{equation}
\noindent
where $I$ is the infinitely large constant depending on the number, masses, and
spins of quantum fields (see the pedagogical derivation of this equation by the
method of dimensional regularization in \cite{28}). The structure of {Equation} (\ref{eq19})
is
the same as the cosmological term in {Einstein's} equation (\ref{eq3}).

The {only}
difficulty is that the constant $I$ is diverging. By making the cutoff
at the Planck momentum $p_{\rm Pl}=(\hbar c^3/G)^{1/2}$, one obtains the
enormously large value of $I\approx 2\times 10^{68}~\mbox{m}^{-2}${ which} exceeds
the observed cosmological constant $\Lambda$ by 120 orders of magnitude \cite{29,30}.
This discrepancy was called the {``vacuum catastrophe''} \cite{31}.

One can argue, however, that the large value of $I$ is determined by the contribution
of virtual particles and, thus, is of no immediate physical meaning. The energy
density $c^4I/(8\pi G)${ determined} by the constant {$I$} does not gravitate as
 the {bare (non-renormalized)}
 electric
 charge in {quantum electrodynamics}
is not a
source of the measurable electric field. This is also in some analogy to the
Casimir effect \cite{32}{ where} only a difference between two infinite energy densities
in the presence and in the absence of plates is the source of gravitational
interaction \cite{33,34} and gives rise to the measurable force.
The quantity $I$ takes the
{physical (renormalized)}
value of
$\Lambda\approx 2\times 10^{-52}~\mbox{m}^{-2}$
only after the renormalization procedure.
These considerations have already received some substantiation in the framework
of quantum field theory in curved space-time, but could obtain the fully
rigorous justification only after a construction of the quantum theory of
gravitation.

There {is}
a lot of different approaches to the problem of cosmological
constants in connection with the origin and physical nature of dark energy.
The number of articles devoted to this subject is large and we do not
aim to review them here (see \cite{37} for an introduction to the field).

\section{Discussion}\label{discuss}

In previous
{Sections,}
{we}
reviewed the main scientific concepts
regarding the structure of the universe developed in the history
of mankind from Ptolemy to Einstein. All of  these concepts imply
that the universe is stationary and its properties do not vary
with time. Although Copernicus and Newton's pictures of the
universe are significantly different from Ptolemy's picture,
all of
{these pictures}
are similar in one basic point: {each
is} time-invariant. This characteristic point was
remained untouched by {Einstein,} who has had in his mind
the predetermined
aim to obtain the stationary cosmological model in the
framework of the general theory of {relativity} developed by him. Actually,
the mathematical formalism of this theory presumed a much more
broad spectrum of cosmological {models} describing the expanding
universe. However, the power of tradition was so strong that
even a great innovator like Einstein chose to modify his
equations by introducing the cosmological term for the sole
purpose of keeping the universe stationary.

This gives us an insight into the {fundamental}
importance of the
scientific results obtained by Alexander Friedmann 100
years ago. By solving the same equations of the general
relativity theory as Einstein, Alexander Friedmann demonstrated
that their general cosmological solution describes the expanding
universe whereas the stationary solution is only a particular
case. This result was so unexpected that Einstein rejected it
as a mathematical error, and only after a detailed written
{explanations} passed to {him}
by Friedmann, {Einstein}  {had}
to
recognize
{that, in fact,}
he himself made an error.

{We} considered the main facts of Alexander Friedmann's scientific
{biography} which provides an explanation why a {mathematician},
educated in the traditions of Saint Petersburg mathematical {school,}
was able to make {such a fundamental}
discovery in the field of
theoretical physics. After the brief exposition of the properties
of expanding universe based on the Friedmann equations, {we}
discussed the main experimental confirmations of the universe
expansion, i.e., the discoveries of redshifts in the spectra of
remote galaxies and the relic~radiation.

Although the theory of the hot universe raised the possibility to
describe the main stages of its evolution, the problem
of cosmological singularity and the problem of horizon remained
unsolved. The solution of these {problems} suggested by the theory
of cosmic inflation links {them}
 to the concept of the quantum
vacuum. There are reasons to believe that the inflationary stage
of the universe expansion is caused by the vacuum quantum effects
of fields of matter rather than by some special inflaton field.
This point of view may find confirmation in further developments
of the quantum theory of gravitation, on the one hand, and by
measurements of the spectrum of relic gravitational radiation,
on the other~hand.

The discovery of the acceleration in the universe expansion
resumed an interest to the cosmological term in the Einstein
equations originally introduced with a single aim to make the
universe stationary. The point is that this term has the same
geometrical form as the vacuum stress-energy tensor of
quantized matter fields and provides a possible explanation of
the observed acceleration of the universe expansion with some
definite value of the cosmological constant. There is a great
discrepancy between this value and theoretical predictions of
quantum field theory which created a discussion in the {literature.}
However, independently of the resolution of this issue, one can
argue that the quantum vacuum bears a direct relation to both
the earliest and modern stages of the evolution of our universe.

\section{Conclusions}\label{concl}

To conclude, Alexander Friedmann made a prediction of the universe
{expansion} which radically altered our scientific picture of the
world as compared to the previous epochs. Later, this prediction
was confirmed experimentally, and {Friedmann}
 equations became the
basis of modern cosmology. Because of this, the Friedmann name
should be put on a par with the names of Ptolemy and Copernicus,
who created the previous, stationary, pictures of the world
around us.

Friedmann's discovery was based on the classical general relativity
theory and could not take into account the quantum effects.
{Nowadays,}
{we}
know that the quantum vacuum plays an important role in the problem
of the origin of Friedmann's universe from the initial singularity,
governs the process of cosmic inflation, and can be considered as
a possible explanation of the observed acceleration of the universe
expansion. Future development of quantum gravity will make our
current knowledge more complete but the concept of expanding
{universe} created by Alexander {Friedmann} will forever remain the
cornerstone of our picture of the world.
}
\vspace{6pt}
\authorcontributions{{Investigation, both authors; writing, both authors. All authors have read and agreed to the published version of the manuscript. }}
\funding{This
 work
was supported by the Peter the Great
Saint Petersburg Polytechnic
University in the framework of the Russian State Assignment for Basic Research
(Project No. FSEG-2020-0024).
{The} work of V.M.M. was supported by the Kazan Federal University
Strategic Academic Leadership Program.
}


\conflictsofinterest{The authors declare no conflict of interest.}


\reftitle{References}




\end{paracol}
\end{document}